\documentclass[lettersize,journal]{IEEEtran}
\usepackage{amsmath,amsfonts,amssymb}
\usepackage{algorithmic}
\usepackage{algorithm}
\usepackage{array}
\newcolumntype{L}[1]{>{\raggedright\arraybackslash}p{#1}}
\newcolumntype{C}[1]{>{\centering\arraybackslash}p{#1}}
\usepackage{booktabs}
\usepackage{makecell}
\usepackage{amsmath}
\usepackage{multirow}
\usepackage[caption=false,font=normalsize,labelfont=sf,textfont=sf]{subfig}
\usepackage{tabularx}
\usepackage{textcomp}
\usepackage{stfloats}
\usepackage{url}
\usepackage{verbatim}
\usepackage{graphicx}
\usepackage{cite}
\begin{document}

\title{TDMOpt: LLM-Driven Cross-layer Dispatch for Transmission-Distribution-Microgrid Systems}

\author{Jiayang Tu, Bojun Du,~\IEEEmembership{Student Member,~IEEE}, Jijun Shui,\\Ershun Du,~\IEEEmembership{Member,~IEEE,} and Qin Wang,~\IEEEmembership{Senior Member,~IEEE}
\thanks{Jiayang Tu, Jijun Shui and Qin Wang are with the Department of Electrical and Electronic Engineering, The Hong Kong Polytechnic University, Hong Kong.}
\thanks{Bojun Du is with the Department of Electrical Engineering, Tsinghua University, Beijing 100084, China}
\thanks{Ershun Du is with the Laboratory of Low Carbon Energy, Tsinghua University, Beijing 100084, China.}
}

\maketitle

\begin{abstract}
Coordinating transmission networks, distribution networks, and microgrids enables distributed flexibility to support renewable energy integration and low-carbon operation.
However, heterogeneous resources, cross-layer coupling, and varying operating requirements complicate the selection of models and coordination schemes that are both physically adequate and computationally tractable.
This paper proposes TDMOpt, a LLM-driven multi-agent framework that replaces fixed modeling configurations with adaptive model composition.
A layered component library provides executable network, resource, objective, constraint, and coordination models.
Guided by retrieved physics and component knowledge, the agents infer specific modeling requirements and jointly select compatible components across layers according to operating conditions and computational budgets.
The model compositions undergo independent AC power flow validation, since feasibility in a simplified formulation does not ensure compliance with physical network limits.
A refinement mechanism interprets numerical feedback to identify modeling requirements for reconsideration and revise relevant component selections.
Case studies on different scale systems illustrate the framework's ability to adapt model combinations and obtain validated dispatch solutions through targeted refinement under varying operating conditions.
\end{abstract}
  

\section{Introduction}
\IEEEPARstart{T}{he} increasing penetration of variable renewable generation and distributed energy resources has shifted substantial flexibility to distribution networks and microgrids~\cite{li2024renewableoperation}.
Its system-level utilization is constrained by local network conditions and interface capabilities~\cite{singhal2023var}, requiring coordinated operation across network levels~\cite{ruwaida2023coordination}.
Low-carbon dispatch further couples power allocation with carbon objectives and constraints~\cite{sang2024carbonflow}.
Hence, coordinated transmission--distribution--microgrid dispatch depends on interdependent network, resource, objective, constraint, and coordination models. 
As operating conditions and task requirements vary, so can the appropriate model combination.

Research on coordinated dispatch has developed complementary models for cross-network coordination, local operating limits, and carbon-related decisions.
At the transmission--distribution interface, market integration and coordination procedures incorporate network limitations into flexibility trading for congestion management~\cite{attar2024marketintegration}.
On the distribution side, operating envelopes derived from linearized three-phase optimal power flow (OPF) translate network constraints into time-varying customer import and export limits~\cite{liu2022operatingenvelope}.
For distribution networks and multiple microgrids, decentralized low-carbon operation has further coupled carbon responsibility allocation with electricity pricing~\cite{ye2025lowcarbon}.
Complementing these operational models, PowerModelsITD.jl supports the construction and comparison of different transmission--distribution OPF formulation combinations within a common optimization platform~\cite{ospina2024powermodelsitd}.
These developments provide modeling approaches and implementation tools on which task-dependent model composition can build.

Selecting among the available formulations requires considering both physical adequacy and computational burden.
Benchmarking of distribution OPF formulations shows that linear approximations, convex relaxations, and nonlinear models differ in feasibility and computational performance~\cite{jha2023dopf}.
To accommodate changing operating conditions, modular linear power flow modeling uses clustered historical scenarios to select approximations from operating regions~\cite{long2024modular}.
Sequential linear approximations have also been used to improve the tractability of stochastic multi-period AC security-constrained OPF~\cite{alizadeh2023tractable}.
Overall, these studies illustrate efforts to balance modeling accuracy and computational tractability.

However, network fidelity is only one aspect of dispatch model.
Operational rules for transformer taps and reactive control devices can require explicit discrete and complementarity constraints within an AC OPF formulation~\cite{alencar2024discrete}.
For storage, exact relaxation of charging--discharging complementarity can depend on locational marginal prices and loss-penalty terms, linking resource constraints to the objective formulation~\cite{wang2024storagerelaxation}.
Carbon-aware OPF additionally incorporates carbon-emission-flow equations and storage carbon-footprint models into the optimization formulation~\cite{chen2025carbonaware}.
Across network levels, second-order cone programming (SOCP)-based transmission--distribution co-optimization combines inter-network data exchange with distributed optimization within distribution networks~\cite{chowdhury2025tdopf}.
Thus, selecting a network formulation alone does not determine a complete configuration of resource, coordination, objective, and constraint components.

These requirements turn task-dependent model composition into a reasoning problem over natural-language intent, domain knowledge, and executable optimization structures.
Recently, large language models (LLMs) have been applied to interpreting operating requirements and supporting optimization and simulation tasks~\cite{yan2024linguistic,bernier2025powergraph,jia2025llmsimulation}.
LLM-based optimization agents can translate natural-language problem descriptions into mathematical formulations and code, and iteratively revise them using execution feedback~\cite{ahmaditeshnizi2024optimus}.
Retrieval augmented generation (RAG) can further ground language models in external knowledge and improve factuality beyond parametric knowledge alone~\cite{asai2024selfrag}.
In power systems, multi-agent LLM frameworks have been used to extract dispatch information, formulate optimization problems, and generate executable code~\cite{yang2026llmdispatch}.
Nevertheless, studies of classical planning show that LLM-generated plans may be non-executable and can benefit from external verification~\cite{valmeekam2023planning}.
These advances provide tools for assisting dispatch modeling, while numerical solvers and independent validation remain necessary for assessing the resulting solutions.

Building on these capabilities, task-adaptive model composition offers a way to retain the physical relationships required by each task without imposing unnecessary model complexity.
However, the availability of modeling and feedback tools does not by itself establish how components with different approximation assumptions and cross-layer dependencies should be jointly selected and revised.
This raises three interrelated challenges for coordinated dispatch.
First, task requirements must be linked to the actual equations of executable components.
Some constraints may be present in several approximations, but their presence does not establish whether the resulting schedule will pass an independent AC check under the current operating condition~\cite{chen2023dispatchable}.
Second, component relevance does not establish joint compatibility. Activated objectives and constraints require variables, equations, and interface quantities that must be supported consistently across the selected components.
A component list therefore requires explicit dependency checks rather than language-level selection alone.
Third, a failed AC check does not uniquely identify the responsible network layer or the appropriate model revision.
Numerical feedback must be interpreted with the current composition and previous trials to guide selective changes within the evaluation budget.
The problem is thus to connect LLM proposals with component physical representations, cross-layer compatibility checks, and validation evidence, rather than to obtain a model recommendation alone.

To address these challenges, this paper proposes an LLM-driven adaptive model composition framework for coordinated transmission--distribution--microgrid low-carbon dispatch.
The framework integrates task interpretation, knowledge retrieval, model composition, and validation-guided refinement.
LLM agents propose and revise component combinations, while registered model builders, numerical solvers, and deterministic checkers instantiate the models, compute dispatch schedules, and assess their validity.
The main contributions are summarized as follows:

\textit{1)} An LLM-driven framework is developed for adaptive model composition in coordinated transmission--distribution--microgrid low-carbon dispatch.
It selects model combinations and coordination schemes according to operating conditions and task requirements, adapting the formulation rather than only its parameters.

\textit{2)} A layered executable component library is constructed with structured descriptions of equations, approximation assumptions, and interfaces.
Joint selection accounts for dependencies across layers, while deterministic checks reject combinations that omit mandatory functions or violate predefined compatibility conditions.

\textit{3)} A refinement mechanism is designed to distinguish execution issues from modeling deficiencies using numerical feedback and the current composition.
It guides selective component revisions, validates the revised candidates, and retains accepted candidates during further exploration.

\section{Problem Formulation}
\subsection{Coordinated Low-Carbon Dispatch Problem}
Consider a coordinated transmission--distribution--microgrid (TDM) system operated over a scheduling horizon $\mathcal{T}$.
The TN, DNs, and MGs use distinct network and resource models coupled through constraints and coordination mechanisms.
Alternative formulations differ in physical fidelity and computational burden, while coordination schemes differ in convergence and scalability.
The model composition is therefore adapted to operating conditions, task requirements, and computational budgets to retain the required physics without unnecessary complexity.

A dispatch task is initiated by a user instruction, which may specify operating conditions, objective priorities, carbon requirements, and security requirements. 
Accordingly, each task is represented as a task-driven optimization problem assembled from appropriate model components. 
Let $q_s$ denote the user instruction in natural language for task $s$. 
The instruction $q_s$ is converted into a structured task descriptor
\begin{equation}
\boldsymbol{\phi}_s =
\left(
\boldsymbol{\phi}_s^{\mathrm{op}},
\boldsymbol{\phi}_s^{\mathrm{obj}},
\boldsymbol{\phi}_s^{\mathrm{req}},
\boldsymbol{\phi}_s^{\mathrm{bud}}
\right)
\label{eq:task_descriptor}
\end{equation}
where $\boldsymbol{\phi}_s^{\mathrm{op}}$ describes the operating conditions; 
$\boldsymbol{\phi}_s^{\mathrm{obj}}$ describes the dispatch preference; 
$\boldsymbol{\phi}_s^{\mathrm{req}}$ describes the operational requirements; 
and $\boldsymbol{\phi}_s^{\mathrm{bud}}$ describes the computational requirement. The descriptor in \eqref{eq:task_descriptor} is not a fixed scenario template. Instead, it provides a structured representation of the task information required for subsequent model selection.

Given the component set $\mathcal{C}_s$ selected for task $s$, the dispatch model is assembled from the variables, objective terms, and constraints introduced by these components. 
The decision variables are denoted as
\begin{equation}
\boldsymbol{x}_s =
\left(
\boldsymbol{x}_s^{T},
\boldsymbol{x}_s^{D},
\boldsymbol{x}_s^{M},
\boldsymbol{x}_s^{I}
\right)
\label{eq:decision_vector}
\end{equation}
where $\boldsymbol{x}_s^{T}$, $\boldsymbol{x}_s^{D}$, and $\boldsymbol{x}_s^{M}$ are the variables of the TN, DN, and MGs, respectively; 
and $\boldsymbol{x}_s^{I}$ represents the interface variables among different layers. 
All decision variables in optimization depend on the selected model components.

The coordinated low-carbon dispatch problem is written as
\begin{align}
\min_{\boldsymbol{x}_s} \quad
F_s(\boldsymbol{x}_s)
&=
\sum_{o \in \mathcal{O}_s}
\omega_{s,o} f_o(\boldsymbol{x}_s)
\label{eq:component_objective}
\\
\mathrm{s.t.} \quad
& g_j(\boldsymbol{x}_s)=0,
\quad j\in \mathcal{G}_s
\label{eq:component_equalities}
\\
& q_j(\boldsymbol{x}_s)\leq 0,
\quad j\in \mathcal{Q}_s 
\label{eq:component_inequalities}
\end{align}
where $\mathcal{O}_s$ is the set of objective components activated for task $s$, 
$f_o(\cdot)$ is the corresponding objective term, 
and $\omega_{s,o}$ is its weight or penalty coefficient; 
$\mathcal{G}_s$ and $\mathcal{Q}_s$ are the equality and inequality constraints generated by the selected components. 
The constraint sets $\mathcal{G}_s$ and $\mathcal{Q}_s$ include the constraints of the selected TN, DN, and MG models, the coupling constraints introduced by coordination components, and additional requirements.
Therefore, the objective function is not fixed for all tasks, but according to the dispatch requirements.

The formulation in \eqref{eq:decision_vector}--\eqref{eq:component_inequalities} represents a family of dispatch models assembled from the component library. 
A fixed baseline is a predefined instance of this family, while the proposed method selects $\mathcal{C}_s$ according to the task descriptor $\boldsymbol{\phi}_s$.

\subsection{Multi-Level Model Components}
The component library provides the building blocks used to assemble the dispatch model in \eqref{eq:decision_vector}--\eqref{eq:component_inequalities}. 
It is organized according to the physical hierarchy of the coordinated system and the functional role of each component:
\begin{equation}
\mathcal{C}=
\mathcal{C}^{T}
\cup
\mathcal{C}^{D}
\cup
\mathcal{C}^{M}
\cup
\mathcal{C}^{C}
\cup
\mathcal{C}^{O}
\cup
\mathcal{C}^{H}
\label{eq:component_library}
\end{equation}
where $\mathcal{C}^{T}$, $\mathcal{C}^{D}$, $\mathcal{C}^{M}$, $\mathcal{C}^{C}$, $\mathcal{C}^{O}$, and $\mathcal{C}^{H}$ denote the TN, DN, MG, coordination, objective, and constraint components, respectively. 
For task $s$, the selected component set satisfies $\mathcal{C}_s \subseteq \mathcal{C}$.
Each component is described by
\begin{equation}
c_i=
\left(
\ell_i,
r_i,
\mathcal{X}_i,
\mathcal{F}_i,
\mathcal{G}_i,
\mathcal{Q}_i,
\rho_i
\right)
\label{eq:component_definition}
\end{equation}
where $\ell_i$ is the layer of component $c_i$, $r_i$ is its ability; 
$\mathcal{X}_i$ denotes the variables introduced by the component; 
$\mathcal{F}_i$ denotes its objective terms; 
$\mathcal{G}_i$ and $\mathcal{Q}_i$ denote its equality and inequality constraints. 
The term $\rho_i$ stores meta information, such as model fidelity, computational burden, and compatibility conditions. 
A component may contribute only part of these elements. 

\begin{table}[h]
\caption{Layered Component Library}
\label{tab:component_library}
\centering
\footnotesize
\renewcommand{\arraystretch}{1.5}
\setlength{\tabcolsep}{4pt}
\begin{tabular}{@{}L{0.20\columnwidth} L{0.74\columnwidth}@{}}
\toprule
Layer & Components \\
\midrule
TN & DC OPF, linearized AC OPF, SOCP AC OPF, AC OPF \\
DN & LinDistFlow, fixed-loss LinDistFlow, SOCP DistFlow, piecewise-linear SOCP DistFlow, DistFlow \\
MG & GT ramping, continuous GT commitment relaxation, BESS preconditioning, flexible-load peak shaving, active balance, bess soc model, renewable curtailment, GT dispatch, carbon model \\
Coordination & centralized, sequential, hierarchical, ADMM, price-based\\
Objective & voltage deviation, T–D interface usage, GT ramping, network-loss penalty, fuel cost, carbon cost, renewable curtailment, load shedding, BESS throughout, MG comfort loss \\
Constraints & TN constraints, DN constraints, MG constraints, carbon intensity limit, reserve margin, ramp limit, system power balance, load shedding limit, boudary power consistency\\
\bottomrule
\end{tabular}
\end{table}

The component library contains listed component types across the six layers in Table~\ref{tab:component_library}.
The mathematical formulations, modeling assumptions, and executable implementations are available in \cite{TDMOptGitHub}.
For the TN and DN, one base formulation is selected to define the main network model. 
MG, objective, and constraint components can be added according to the task descriptor. 
Coordination components define the coupling and solution scheme among the TN, DN, and MGs. 
The component library defines the available model components, while the selection of $\mathcal{C}_s$ for a given task is formulated as an adaptive model composition problem in the section II.C.

\subsection{Adaptive Model Composition Problem}
Based on the component library, a dispatch formulation for task $s$ is determined by selecting components and associated parameters. 
Let a model composition be denoted as
\begin{equation}
\boldsymbol{m}_s =
\left(
\mathcal{C}_s,
\boldsymbol{\theta}_s
\right),
\quad
\mathcal{C}_s \subseteq \mathcal{C}
\label{eq:model_composition}
\end{equation}
where $\boldsymbol{\theta}_s$ contains the corresponding parameters of the dispatch model. 
Different choices of $\boldsymbol{m}_s$ lead to different dispatch formulations, even under the same physical system and operating data.

Howerver, not all component combinations are valid. 
Therefore, the admissible composition set for task $s$ is defined as
\begin{equation}
\mathcal{A}_s =
\left\{
\boldsymbol{m} \in \mathcal{M}
\mid
\mathcal{R}
\left(
\boldsymbol{m},
\boldsymbol{\phi}_s
\right)
\right\}
\label{eq:admissible_set}
\end{equation}
where $\mathcal{M}$ denotes the model composition space, and
$\mathcal{R}(\boldsymbol{m},\boldsymbol{\phi}_s)$ is a boolean value that holds when the components in $\boldsymbol{m}$ are compatible and satisfy the requirements of task $s$.

For each admissible composition $\boldsymbol{m}_s\in\mathcal{A}_s$, the assembled dispatch model produces a solution $\boldsymbol{x}_s^\star(\boldsymbol{m}_s)$ by solving \eqref{eq:component_objective}--\eqref{eq:component_inequalities}. 
The quality of a composition is evaluated by
\begin{equation}
J_s(\boldsymbol{m}_s)
=
\eta_1 C_s^{\mathrm{op}}
+
\eta_2 E_s^{\mathrm{CO_2}}
+
\eta_3 V_s^{\mathrm{val}}
+
\eta_4 T_s^{\mathrm{sol}}
\label{eq:composition_score}
\end{equation}
where $C_s^{\mathrm{op}}$ is the operation cost, $E_s^{\mathrm{CO_2}}$ is the carbon emission, $V_s^{\mathrm{val}}$ measures the violation after validation, and $T_s^{\mathrm{sol}}$ is the solution time. 
The coefficients $\eta_1$--$\eta_4$ reflect the task preference in $\boldsymbol{\phi}_s$.

The desired model composition can be written as
\begin{equation}
\boldsymbol{m}_s^\star
=
\arg\min_{\boldsymbol{m}_s\in\mathcal{A}_s}
J_s(\boldsymbol{m}_s)
\label{eq:adaptive_composition_problem}
\end{equation}
However, directly solving \eqref{eq:adaptive_composition_problem} by exhaustive enumeration is inefficient because the number of possible compositions grows rapidly with the number of network models, coordination schemes, objective terms, and additional constraints. 
A typical search space can be on the order of
$|\mathcal{A}|=\mathcal{O}\left(|\mathcal{C}^{T}||\mathcal{C}^{D}||\mathcal{C}^{C}|2^{|\mathcal{C}^{M}|+|\mathcal{C}^{O}|+|\mathcal{C}^{H}|}\right)$.
Therefore, the main challenge is not only to find a feasible dispatch solution, but also to identify a suitable model composition with a small number of trials. 
This motivates the proposed LLM-driven adaptive model composition in the next section.

\section{Proposed Method}
The adaptive composition problem in \eqref{eq:adaptive_composition_problem} requires identifying appropriate models from a large and heterogeneous composition space. 
Exhaustive evaluation is impractical because the number of candidate compositions grows rapidly and each candidate must be assembled, solved, and validated. 
Fixed selection rules are also difficult to generalize across changing operating conditions, modeling requirements and task requirements.
To address this challenge, this paper proposes an LLM-driven adaptive model composition framework that integrates task interpretation, component retrieval, model composition, and verification-guided refinement into a closed-loop workflow. 
\subsection{LLM-Driven Adaptive Model Composition Framework}
\begin{figure}[!t]
\centering
\includegraphics[width=\columnwidth]{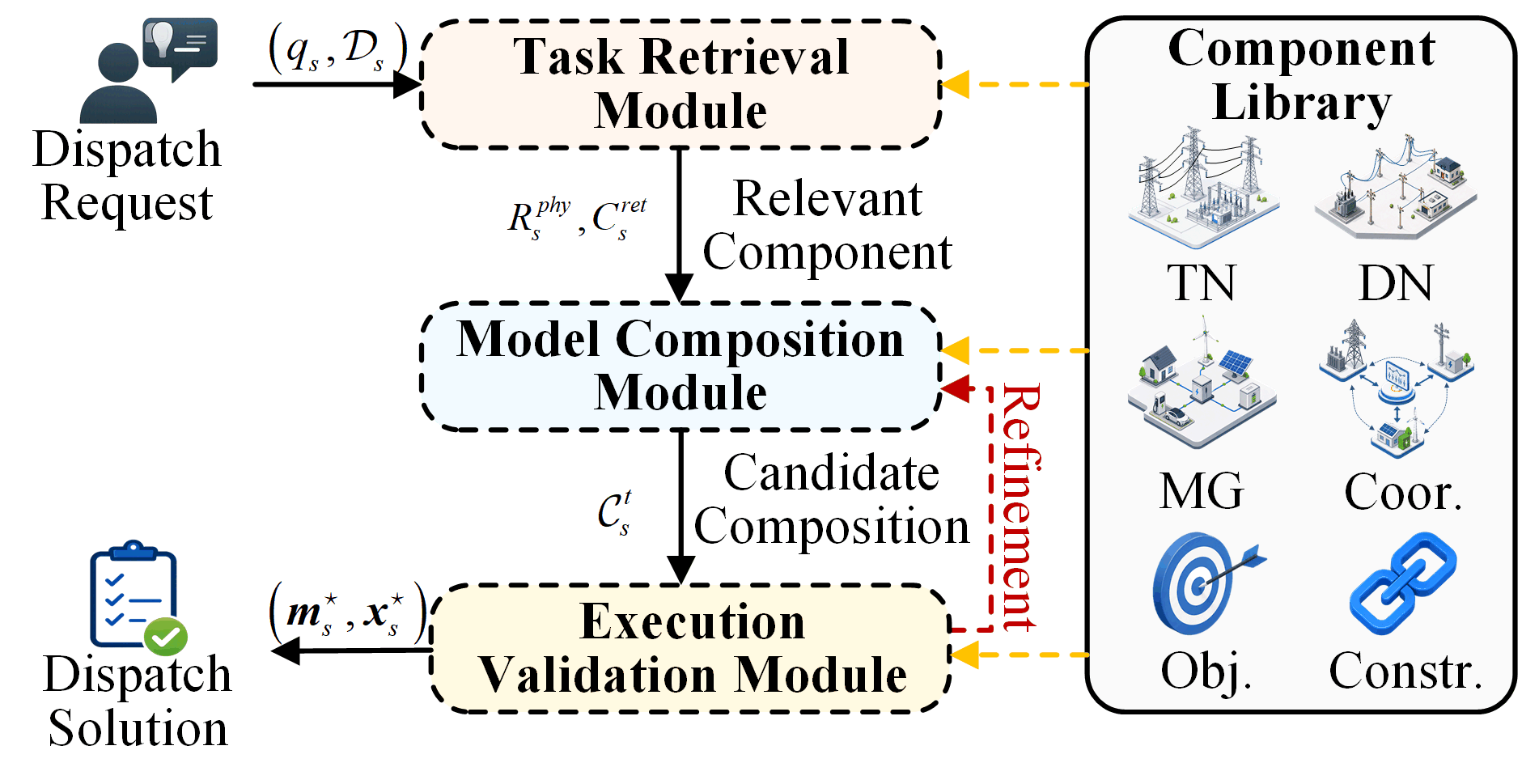}
\caption{Overall framework of the proposed LLM-driven adaptive model composition method.}
\label{fig:framework}
\end{figure}

\begin{figure*}[!t]
\centering
\includegraphics[width=\textwidth]{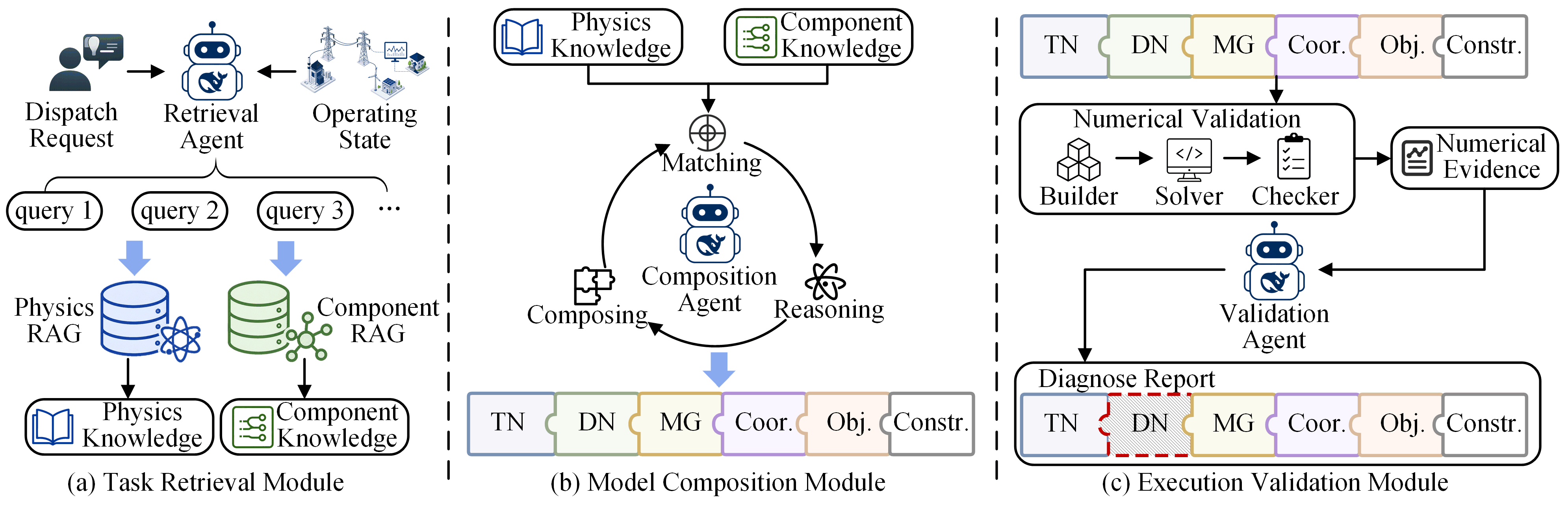}
\caption{Distinct reasoning roles of the three LLM agents in the proposed framework: 
(a) retrieval of physics knowledge and inference of modeling requirements, 
(b) model composition using retrieved knowledge, and 
(c) diagnosis and refinement using numerical validation results.}
\label{fig:agent_roles}
\end{figure*}

Fig.~\ref{fig:framework} presents the overall framework, which consists of a task retrieval module, a model composition module, and an execution validation module.
The component library provides structured descriptions for retrieval and composition and executable builders for model instantiation.
Fig.~\ref{fig:agent_roles} complements the overall architecture by illustrating the distinct roles of the three LLM agents within the corresponding modules.

Let $\mathcal{D}_s$ denote the system and operating data associated with task $s$. 
Starting from the user instruction $q_s$ and the data $\mathcal{D}_s$, the proposed framework follows the workflow 
\begin{equation}
\left(q_s,\mathcal{D}_s\right)
\longrightarrow
\boldsymbol{\phi}_s
\longrightarrow
\mathcal{C}_s^{\mathrm{ret}}
\longrightarrow
\mathcal{C}_s^{t}
\longrightarrow
\left(
\boldsymbol{m}_s^{t},
\boldsymbol{x}_s^{t},
\boldsymbol{r}_s^{t}
\right)
\label{eq:overall_workflow}
\end{equation}
where $\mathcal{C}_s^{\mathrm{ret}}\subseteq\mathcal{C}$ is the task-relevant component set retrieved from the component knowledge base, and
$\mathcal{C}_s^{t}\subseteq\mathcal{C}_s^{\mathrm{ret}}$ is the component combination generated at iteration $t$.
Within the task retrieval module, the intermediate set $\mathcal{R}_s^{\mathrm{phy}}$ denotes the physical modeling requirements inferred from $\boldsymbol{\phi}_s$ and the retrieved physics knowledge, and is used to guide the retrieval of $\mathcal{C}_s^{\mathrm{ret}}$.
The executable model, dispatch solution, and validation report obtained at this iteration are denoted by $\boldsymbol{m}_s^{t}$, $\boldsymbol{x}_s^{t}$, and $\boldsymbol{r}_s^{t}$, respectively.

The task retrieval module converts the user instruction $q_s$ and operating data $\mathcal{D}_s$ into the structured descriptor $\boldsymbol{\phi}_s$.
Based on these task facts, the retrieval agent formulates queries to retrieve relevant physics and component knowledge.
The retrieved physics knowledge is used to infer $\mathcal{R}_s^{\mathrm{phy}}$, while the component retrieval produces the task-relevant set $\mathcal{C}_s^{\mathrm{ret}}$.
This step identifies the physical relationships that should be represented for the current dispatch task without directly prescribing a specific model formulation.

The model composition module uses retrieved physics knowledge $\mathcal{R}_s^{\mathrm{phy}}$ to match task-relevant component and construct $\mathcal{C}_s^{t}$.
The composition agent relates the required physical relationships to the variables, equations, assumptions, and modeling capabilities of candidate components, while considering their cross-layer dependencies, compatibility conditions, model fidelity, and computational burden.
The TN, DN, MG, coordination, objective, and constraint components are then jointly assembled into a task-specific candidate model.

Registered model builders and numerical solvers produce the dispatch solution $\boldsymbol{x}_s^{t}$, which is subsequently evaluated by deterministic validation to obtain $\boldsymbol{r}_s^{t}$.
When component revision is required, the validation agent interprets $\boldsymbol{r}_s^{t}$ and returns feedback to the model composition module.
An accepted candidate is retained while further combinations may be explored.
When the stopping condition is reached, the retained model and schedule are returned as $\left(\boldsymbol{m}_s^\star,\boldsymbol{x}_s^\star\right)$, if available.

The LLM agents are used for knowledge retrieval, physical requirement inference, model composition reasoning, and interpretation of numerical feedback.
They neither generate optimization equations nor replace the numerical solver or deterministic checker.
All executable models are instantiated from predefined components, and the final dispatch solution is accepted only after deterministic numerical verification.
The retrieval, composition, and validation mechanisms are detailed in the following subsections.

\subsection{Task Retrieval Module}
Searching directly over the complete component library introduces irrelevant alternatives and increases the complexity of model composition. 
As illustrated in Fig.~\ref{fig:agent_roles}(a), the task retrieval module therefore grounds the task information with external physics and component knowledge before model selection.
The retrieval agent does not directly determine a model formulation; instead, it identifies the physical modeling requirements implied by the task and retrieves executable components that may represent these requirements.

The module first performs deterministic eligibility screening to remove components that cannot be instantiated for the current task.
The screening considers data availability, system topology, solver support, and explicit operational requirements.
For each component layer$\ell\in\{T,D,M,C,O,H\}$, let $\mathcal{C}^{\ell}$ denote the set of all registered components in layer $\ell$, 
and let $\widetilde{\mathcal{C}}_s^{\ell}\subseteq\mathcal{C}^{\ell}$ denote the subset that passes the eligibility checks for task $s$.
The complete executable candidate pool is $\widetilde{\mathcal{C}}_s=\bigcup_{\ell\in}\widetilde{\mathcal{C}}_s^{\ell}$.
This screening only determines whether a component can be instantiated for the current task; its relevance to the inferred physical requirements and its compatibility with components from other layers are evaluated subsequently.

Given the structured task descriptor $\boldsymbol{\phi}_s$ and operating data $\mathcal{D}_s$, the retrieval agent first formulates physics queries according to the operating phenomena and task priorities contained in these inputs.
The queries are submitted to the physics RAG, which retrieves relevant physical principles and relationships from the domain knowledge base.
Let $\mathcal{K}_s^{\mathrm{phy}}$ denote the retrieved physics knowledge.
The retrieval agent then infers the physical modeling requirements as
\begin{equation}
\mathcal{R}_s^{\mathrm{phy}}=\operatorname{Infer}_{\mathrm{LLM}}\left(\boldsymbol{\phi}_s,\mathcal{D}_s,\mathcal{K}_s^{\mathrm{phy}}\right)
\label{eq:physical_requirement_inference}
\end{equation}
The elements of $\mathcal{R}_s^{\mathrm{phy}}$ describe model physical relationships that should be represented, such as reactive power balance, voltage and power coupling, network losses, interface consistency, and intertemporal resource coupling.
Importantly, this inference stage does not assign a specific component or model.

Based on $\boldsymbol{\phi}_s$ and $\mathcal{R}_s^{\mathrm{phy}}$, the retrieval agent further formulates structured queries for the component RAG.
Each component record is constructed from the predefined component description in \eqref{eq:component_definition} and contains its variables, governing equations, modeling assumptions, omitted physical effects, interface quantities, mathematical structure, and implementation information.
These records describe what a component represents rather than prescribing the operating scenarios in which it should be selected.
The task-relevant component set is then retrieved from the eligible set as
\begin{equation}
\mathcal{C}_s^{\mathrm{ret}}=\left\{c_i \in \widetilde{\mathcal{C}}_s\mid\eta_{s,i} \geq \tau_s\right\}
\label{eq:component_retrieval}
\end{equation}
where $\eta_{s,i}$ denotes the retrieval relevance score of component $c_i$ with respect to the task descriptor $\boldsymbol{\phi}_s$ and the inferred physical requirements $\mathcal{R}_s^{\mathrm{phy}}$, and $\tau_s$ is the retrieval threshold.

Components explicitly required by the task are retained during retrieval.
The resulting $\mathcal{R}_s^{\mathrm{phy}}$ and $\mathcal{C}_s^{\mathrm{ret}}$ provide the physical requirements and structured component knowledge used by the model composition agent.
Therefore, the retrieval module reduces and grounds the subsequent model space without encoding direct mappings from operating scenarios to model choices.

\subsection{Model Composition Module}
The retrieved set $\mathcal{C}_s^{\mathrm{ret}}$ contains executable components that are relevant to the task, but does not determine how they should be combined.
A valid dispatch formulation not only represent the physical requirements in $\mathcal{R}_s^{\mathrm{phy}}$, but also maintain consistent dependencies across the TN, DN, MG, coordination, objective, and constraint layers.
As illustrated in Fig.~\ref{fig:agent_roles}(b), the composition agent therefore performs physics-model matching, cross-layer reasoning, and candidate composition over the retrieved component knowledge.
Independent selection within each layer may omit required functions or produce incompatible cross-layer structures. 

The LLM receives the task descriptor $\boldsymbol{\phi}_s$, the operating data $\mathcal{D}_s$, the retrieved component set $\mathcal{C}_s^{\mathrm{ret}}$, and the metadata $\rho_i$ of each retrieved component. 
These inputs allow the LLM to relate the dispatch requirements to the capabilities of the available components.

Each component $c_i\in\mathcal{C}_s^{\mathrm{ret}}$ is accompanied by its structured description, including the represented variables and equations, modeling assumptions, omitted physical effects, interface quantities, mathematical structure, and computational characteristics.
Let $\mathcal{P}_i$ denote the set of physical relationships represented by component $c_i$.
The composition agent first compares $\mathcal{R}_s^{\mathrm{phy}}$ with these component descriptions and identifies combinations whose represented physics collectively cover the task requirements.
Accordingly, a candidate combination should satisfy
\begin{equation}
\mathcal{R}_s^{\mathrm{phy}}
\subseteq
\bigcup_{c_i\in\mathcal{C}_s^{t}}
\mathcal{P}_i
\label{eq:physics_coverage}
\end{equation}
This matching is based on the actual model structure rather than predefined mappings between operating scenarios and model choices.
For example, stressed network conditions or strict voltage limits may require formulations that represent voltage and reactive power, whereas a tight computational budget may favor simplified network models. 
When multiple formulations satisfy the required physics, their approximation assumptions, omitted effects, model fidelity, computational burden, and the task priorities in $\boldsymbol{\phi}_s$ are further considered.

Physics coverage alone is insufficient because individually suitable components may form an incomplete or incompatible model when combined across layers.
The composition agent therefore reasons jointly over the dependencies among retrieved components, including exchanged interface quantities, required state variables, coordination requirements, and dependencies introduced by objectives and constraints.
At iteration $t$, the resulting component combination is represented as
\begin{equation}
\mathcal{C}_s^{t}
=
\left(
\mathcal{C}_s^{T,t},
\mathcal{C}_s^{D,t},
\mathcal{C}_s^{M,t},
\mathcal{C}_s^{C,t},
\mathcal{C}_s^{O,t},
\mathcal{C}_s^{H,t}
\right)
\label{eq:candidate_component_combination}
\end{equation}
where $\mathcal{C}_s^{\ell,t}\subseteq\mathcal{C}_s^{\mathrm{ret}}\cap\mathcal{C}^{\ell}$ denotes the components selected from layer $\ell\in\{T,D,M,C,O,H\}$.
The TN, DN, and coordination layers each contain one component, whereas the MG, objective, and constraint layers may contain multiple components.

The specific layer selections are determined jointly rather than independently.
For example, a network formulation that introduces reactive power or voltage variables must be considered together with the corresponding interface quantities and resource models, while an objective or operational constraint must be supported by the variables and equations required for its evaluation.
This cross-layer reasoning prevents combinations that individually appear relevant but are incomplete when assembled into a coordinated T-D-M dispatch model.

The composition agent produces $\mathcal{C}_s^{t}$ in a predefined structured format containing only registered component identifiers.
A deterministic parser subsequently checks component existence, mandatory functions, cross-layer compatibility, and duplication with previously evaluated candidates.
A candidate that contains unregistered components, omits required functions, or violates predefined compatibility conditions is rejected before model execution.

For the initial candidate, the composition reasoning is based on $\mathcal{R}_s^{\mathrm{phy}}$, $\mathcal{C}_s^{\mathrm{ret}}$, and the associated component knowledge.
For subsequent candidates, diagnostic feedback from the execution validation module identifies insufficient modeling requirements and the affected component layers.
The composition agent then revises the relevant selections while preserving components that are not implicated by the validation result.
Thus, the search uses validation evidence to guide candidate generation rather than exhaustively enumerating the complete composition space.
The model composition module determines only the component set $\mathcal{C}_s^{t}$.
The corresponding model parameters, numerical formulation, and solver configuration are instantiated by the execution validation module.

\subsection{Execution Validation Module}
The execution validation module converts the component combination $\mathcal{C}_s^{t}$ generated at iteration $t$ into an executable optimization model and evaluates whether the resulting solution satisfies the task requirements.
As illustrated in Fig.~\ref{fig:agent_roles}(c), numerical execution and deterministic checking provide objective evidence, while the validation agent interprets this evidence when the current model is inadequate.

The module first configures the model and solver parameters according to the selected components, the task descriptor, and the operating data:
\begin{equation}
\boldsymbol{\theta}_s^{t}=\operatorname{Configure}\left(\mathcal{C}_s^{t},\boldsymbol{\phi}_s,\mathcal{D}_s\right)
\label{eq:parameter_configuration}
\end{equation}
The parameter vector $\boldsymbol{\theta}_s^{t}$ contains the model coefficients, coordination parameters, and solver settings required for execution.
Parameter configuration follows predefined rules associated with the selected components and does not alter the component combination $\mathcal{C}_s^{t}$.

The component combination $\mathcal{C}_s^{t}$ and the corresponding parameter vector $\boldsymbol{\theta}_s^{t}$ form the executable model $\boldsymbol{m}_s^{t}=(\mathcal{C}_s^{t},\boldsymbol{\theta}_s^{t})$, consistent with the model definition in \eqref{eq:model_composition}.
Registered model builders instantiate the variables, objective terms, and constraints defined by the selected components in accordance with \eqref{eq:component_definition}.
The resulting optimization problem follows the general formulation in \eqref{eq:component_objective}--\eqref{eq:component_inequalities}, and the corresponding numerical solver is invoked to obtain the dispatch solution $\boldsymbol{x}_s^{t}$ and its termination status.

Successful solver termination alone does not guarantee that the obtained schedule is physically valid, particularly when simplified network formulations or approximate coordination models are used.
The solution is therefore evaluated by deterministic validation routines:
\begin{equation}
\boldsymbol{r}_s^{t}
=\operatorname{Validate}
\left(
\boldsymbol{x}_s^{t},
\boldsymbol{m}_s^{t},
\boldsymbol{\phi}_s,
\mathcal{D}_s
\right)
\label{eq:validation_report}
\end{equation}
where $\boldsymbol{r}_s^{t}$ denotes the numerical validation report.
The report records the solver status, constraint residuals, and violations of the requirements specified by $\boldsymbol{\phi}_s$.
Depending on the activated components and task requirements, the validation includes power balance, voltage and branch limits, cross-level interface consistency, storage SOC, carbon limits, reserve requirements, and other active constraints.
A higher fidelity power flow calculation is additionally applied when required to assess schedules obtained from simplified network formulations.

After the candidate satisfies the numerical checks, it is established or updated as the retained candidate. 
If the conditions for continued exploration are still met, other combinations are evaluated. 
Once the stopping condition is reached, the retained accepted candidate is returned.
Otherwise, the validation result is first examined to distinguish execution issues from deficiencies in the selected model structure.
Failures caused by numerical settings or other predefined configurable quantities are handled within the execution module without changing $\mathcal{C}_s^{t}$.
When the failure indicates that the current formulation does not adequately represent the required physics, the numerical evidence in $\boldsymbol{r}_s^{t}$ is provided to the validation agent.

The validation agent interprets the violated quantities together with the physical requirements in $\mathcal{R}_s^{\mathrm{phy}}$ and the current component combination $\mathcal{C}_s^{t}$.
It then identifies which modeling requirements are insufficiently represented and which component layers are implicated by the failure.
The resulting diagnostic feedback specifies the requirements that should be reconsidered while preserving unaffected parts of the current composition whenever possible.
The validation agent does not directly select a replacement component; instead, the diagnostic feedback is returned to the model composition module, where the relevant component selections are revised using the procedure described in Section III-C.

The revised candidate is subsequently instantiated, solved, and validated through the same deterministic execution procedure.
In this way, numerical validation serves not only as an acceptance test but also as grounded evidence for targeted model refinement, preventing the LLM agents from relying solely on their initial reasoning when constructing the final dispatch formulation.

\section{Case Study}
\subsection{Experimental Setup}
\begin{figure}[h]
\centering
\includegraphics[width=\columnwidth]{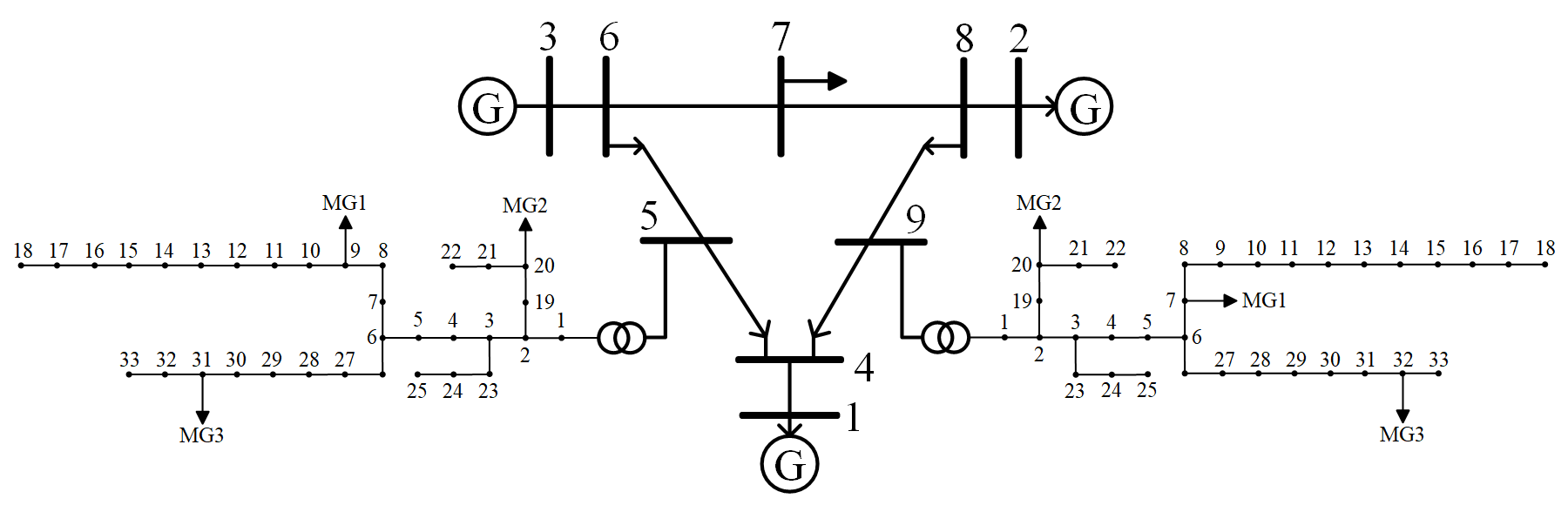}
\caption{Topology of the T9D2 system.}
\label{fig:topology}
\end{figure}

The proposed framework is evaluated on two modified TDM systems, T9D2 and T118D10.
The T9D2 system consists of an IEEE 9-bus TN, two modified IEEE 33-bus DNs, and six MGs, as shown in Fig.~\ref{fig:topology}.
The two DNs are connected to TN buses 5 and 9, respectively, and three thermal units are located at TN buses 1, 2, and 3.
The MGs connected to DN1 are located at buses 9, 20, and 31, whereas those connected to DN2 are located at buses 7, 20, and 32. 
The hierarchical TDM interconnection structure is adapted from the T9D2 configuration in~\cite{Zhai2026TDM}. 
The larger T118D10 system comprises an IEEE 118-bus TN, ten modified 33-bus DNs, and thirty MGs.
Both systems include photovoltaic and wind generation, gas turbines, battery energy storage systems, static var compensators, and flexible loads on the distribution and microgrid sides.
Each MG contains local generation, battery storage, and demand-side resources.
The placement, resource composition, and nominal parameters of distribution-side DERs and MGs in T9D2 are primarily adapted from~\cite{Hao2025FAPVE}.
Selected capacities and operating parameters are adjusted to match the power bases, system scale, and operating scenarios adopted in this study.
For T118D10, the TN generation and network parameters are obtained from PGLib-OPF~\cite{PGLibOPF2307}, while the DN line and nominal load parameters are based
on the 33-bus feeder data in~\cite{Baran1989}.
The DNs are represented as 35-kV regional equivalents, with electrical parameters and resource capacities scaled to their assigned demand levels.
The TN generator emission factors in T118D10 are assigned according to the fuel-type data in~\cite{PGLibCO2v2}.

The hourly load, PV generation, and wind generation profiles are derived from the German historical records provided by Open Power System Data~\cite{OPSD2019,OPSDTimeSeries}.
The selected period extends from January~1 to January~30, 2019, comprising 720 consecutive hourly records.
Each series is normalized by its maximum over the selected period and scaled to the corresponding nominal loads or installed renewable capacities.
Feeder renewable profile adjustments and additional daily operating conditions, including renewable scaling and reactive support availability, are specified for the test systems.
The resulting profiles are used as deterministic day-ahead scheduling inputs. 
Each date defines an independent dispatch instance with a 24h horizon and a 1h resolution.

The proposed workflow is compared with a fixed simplified baseline on both test systems.
The baseline method combines DC OPF for the TN, LinDistFlow for the DNs, and centralized coordination.
The computational comparison for T9D2 additionally includes a fixed full network reference using AC OPF for the TN and full DistFlow for the DNs.
The proposed workflow selects and refines model components according to the operating task and numerical validation feedback.
Each dispatch instance is described by a natural language operating task $q_s$ and the corresponding numerical data $\mathcal{D}_s$.
The user task specifies the operating objectives and requirements, while $\mathcal{D}_s$ provides the network, resource, and hourly operating data.
For each instance, the compared methods use the same resource data and basic operating requirements.
The final schedules are evaluated using the same AC power flow checker, with the admissible voltage range set to 0.95--1.05~p.u.
The terminal state of charge of each battery is required to recover its initial value.
A separate comparison evaluates the initial model compositions produced by Claude Code (CC) and TDMOpt using the same DeepSeek model.

The reported cost and carbon emissions are aggregated over the 30 daily dispatch instances for each test system.
The reported solver time is averaged over the final daily models and excludes LLM search, model assembly, and AC replay.
The framework is implemented in Python 3.11 using Pyomo 6.10. 
The LLM agent uses deepseek-v4-pro, with a maximum output length of 8192 tokens.
All experiments are conducted on a workstation equipped with one Intel Xeon w7-2475X processor and 64~GiB RAM.

\subsection{Overall Dispatch Performance}
Table~\ref{tab:cross_topology_performance} reports composition validity and final dispatch performance for the 30 daily tasks on each system.
For the composition comparison, CC and TDMOpt use the same task inputs and component library.
Success rate is the percentage of provided model candidate that retain mandatory functions, satisfy compatibility conditions, and can be instantiated by the common model builder.
This comparison excludes numerical optimization and AC validation.
CC achieves success rates of 26.7\% on T9D2 and 10.0\% on T118D10, compared with 100\% for TDMOpt on both systems.
For both T9D2 and T118D10, the proposed workflow produces dispatch schedules without AC voltage violations, whereas the fixed baseline exhibits bus voltage violations.
This improvement is accompanied by a modest increase in total operating cost and a higher average solver time, while total carbon emissions are higher or lower than those of the baseline.
These results support the effectiveness of the task-dependent model compositions in providing secure dispatch across the tested daily tasks without high cost or solver time increasing.

\begin{table}[!t]
\caption{Composition Validity and Dispatch Performance}
\label{tab:cross_topology_performance}
\centering
\scriptsize
\renewcommand{\arraystretch}{1.15}
\setlength{\tabcolsep}{2pt}
\begin{tabular*}{\columnwidth}
{@{\extracolsep{\fill}}llccccc@{}}
\toprule
Topology & Method
& \shortstack{Success\\rate (\%)}
& \shortstack{Cost\\($10^6$ \$)}
& \shortstack{Violations\\(bus-hours)}
& \shortstack{Solver\\time (s)}
& \shortstack{Emissions\\($10^3$ t)} \\
\midrule
\multirow{3}{*}{T9D2}
& Baseline
& -- & 14.18 & 563 (1.04\%) & 36.17 & 151.38 \\
& CC
& 26.7\% & -- & -- & -- & -- \\
& Proposed
& 100\% & 14.52 & 0 & 37.44 & 149.42 \\
\midrule
\multirow{3}{*}{T118D10}
& Baseline
& -- & 65.15 & 10255 (3.41\%) & 776.22 & 1600.79 \\
& CC
& 10\% & -- & -- & -- & -- \\
& Proposed
& 100\% & 67.74 & 0 & 1410.76 & 1644.41 \\
\bottomrule
\end{tabular*}
\par\vspace{2pt}
\begin{minipage}{\columnwidth}
\scriptsize
\end{minipage}
\end{table}

\begin{figure}[h]
\centering
\includegraphics[width=\columnwidth]{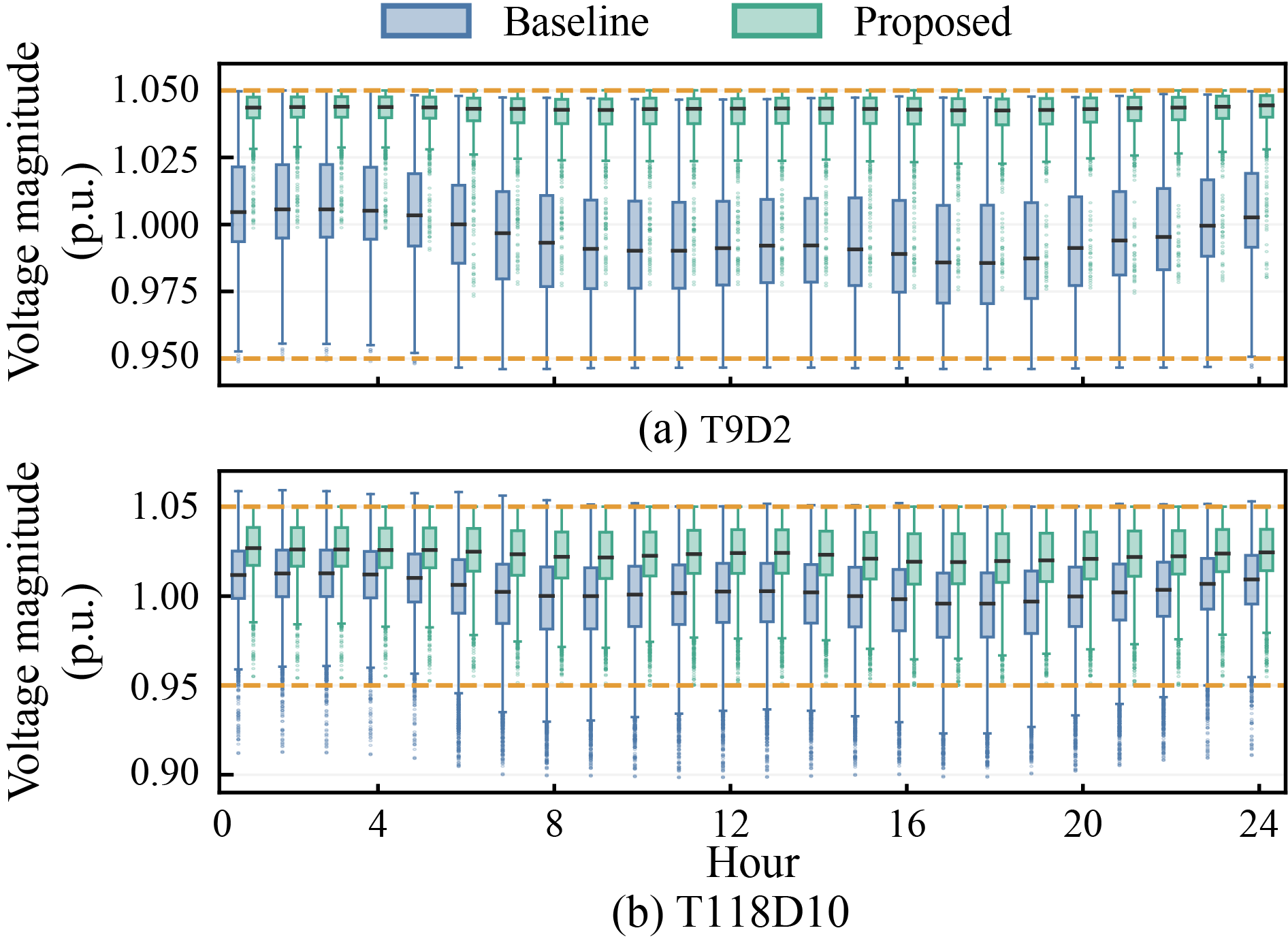}
\caption{Hourly voltage distributions obtained by the baseline and proposed workflows over 30 days of two cases. (a) T9D2 system; (b) T118D10 system.}
\label{fig:Voltage}
\end{figure}

To examine the hourly patterns and magnitudes of the voltage excursions summarized in Table~\ref{tab:cross_topology_performance}, Fig.~\ref{fig:Voltage} compares the hourly voltage distributions over the 30 daily tasks.
For both systems, the baseline distributions extend below the lower voltage limit; the T118D10 baseline also exceeds the upper limit in some hourly groups.
The proposed distributions remain within the displayed limits in both systems.

\begin{figure}[h]
\centering
\includegraphics[width=\columnwidth]{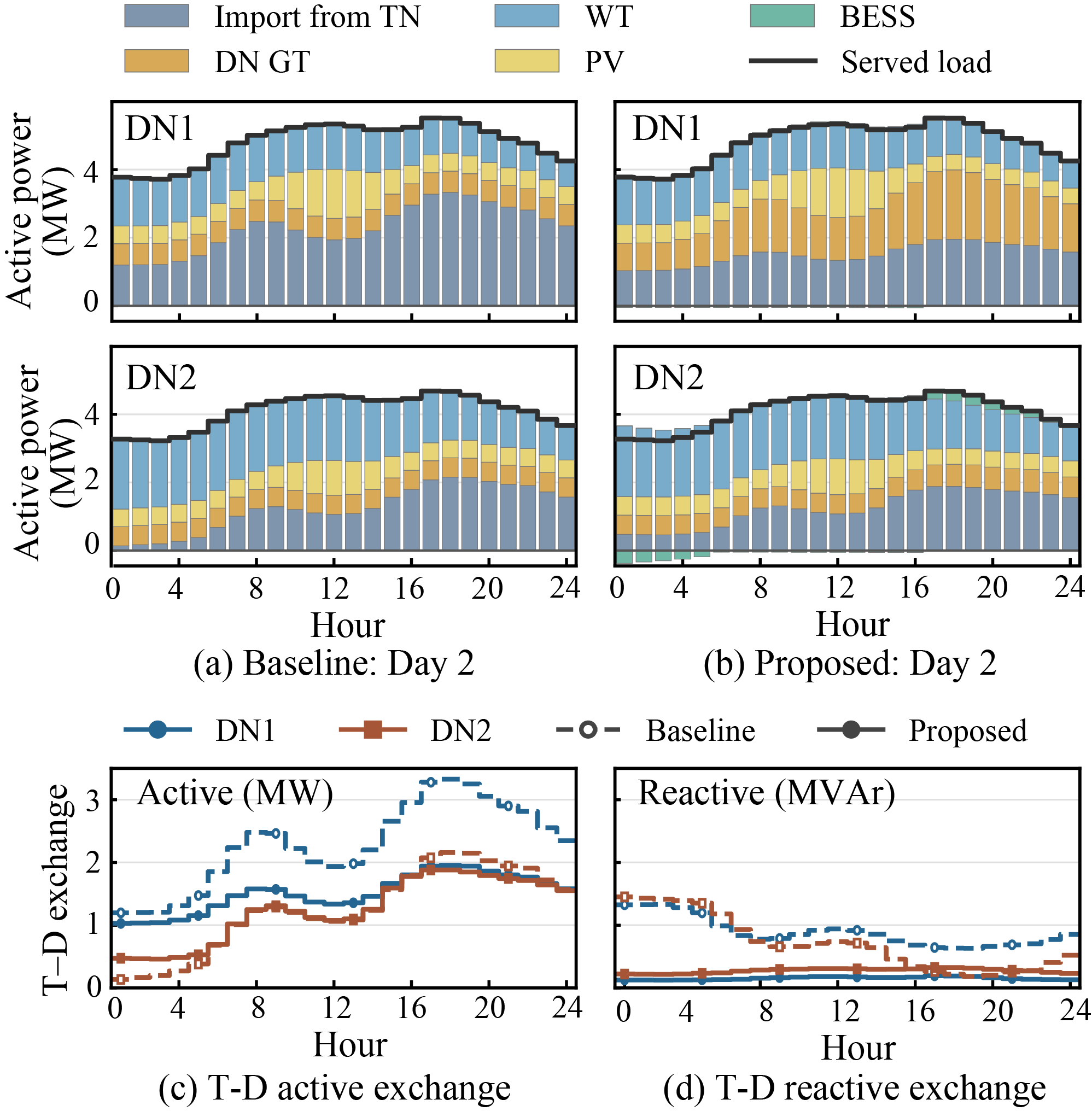}
\caption{Representative day dispatch and T--D power exchanges on T9D2.
(a) Active power dispatch in DN1 and DN2 with its connected MGs under the baseline method;
(b) Active power dispatch in DN1 and DN2 with its connected MGs under the proposed method;
(c) Active power exchanges at the two T--D interfaces;
(d) Reactive power exchanges at the two T--D interfaces.}
\label{fig:Dispatch}
\end{figure}

To examine the operating adjustments associated with the voltage-compliant schedules, Fig.~\ref{fig:Dispatch} compares local dispatch and T--D power exchanges on T9D2 Day~2.
Higher GT output in DN1 coincides with lower active power imports from the TN, whereas DN2 imports increase during the early hours.
Reactive power exchanges at both T--D interfaces are lower over most of the day.
These changes show that the adopted schedule redistributes local supply and T--D exchanges differently in the two DNs.

The improvement in voltage compliance is accompanied by higher solver times relative to the simplified baseline.
This motivates examining whether full TN and DN formulations are needed for every tested task and how the adopted compositions compare computationally with a fixed full reference.

\subsection{Scenario-Adaptive Model Composition}

\begin{figure}[h]
\centering
\includegraphics[width=\columnwidth]{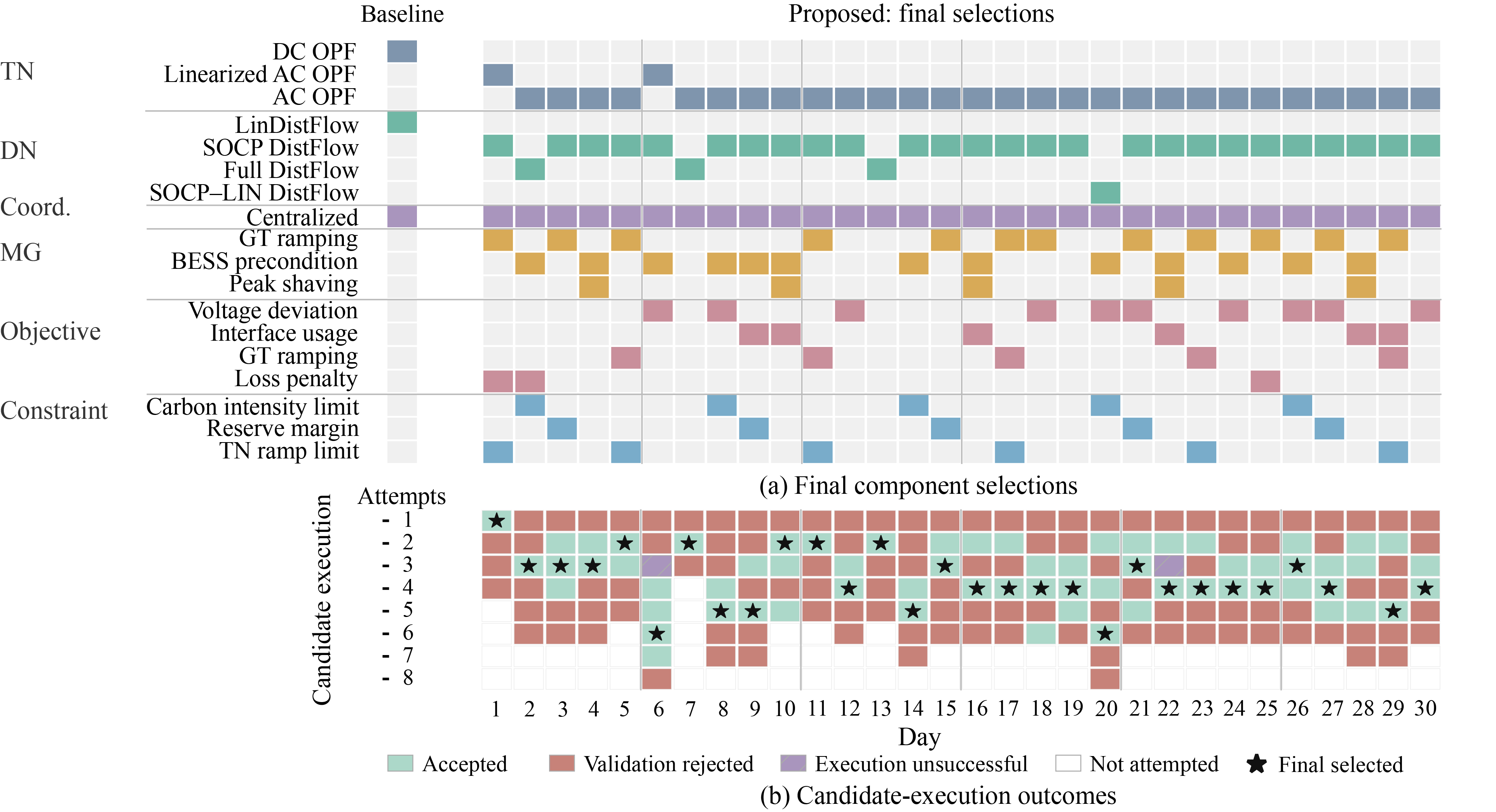}
\caption{Daily model compositions and candidate-execution outcomes on T9D2. 
(a) Component selections in the final adopted formulations over the 30 days; 
(b) Outcomes of successive candidate executions for each day.}
\label{fig:Compositions}
\end{figure}

The schedules reported above are obtained using different model compositions across the daily tasks. 
The analysis focuses on T9D2 to examine the network representations and operating functions retained in the final daily formulations.
In Fig.~\ref{fig:Compositions}(a), AC OPF is used for the TN on most days, whereas SOCP DistFlow predominates for the DNs.
Full DistFlow is retained on Days~2,~7, and~13, and SOCP--LIN DistFlow is selected on Day~20.
These accepted combinations show that the applied validation requirements can be satisfied using different network representations across layers, without imposing the full formulation on every daily task.

This flexibility also concerns the operating functions represented alongside the network equations.
Even when the TN--DN model pair remains unchanged, the selected MG, objective, and constraint components can differ.
For example, Days~8 and~9 both use AC OPF and SOCP DistFlow, but Day~8 includes a voltage deviation objective term and a carbon intensity limit, while Day~9 includes an interface usage objective term and a reserve margin constraint.
Selecting the network-model pair therefore constitutes only part of the composition task: the assembled formulation also specifies the operating functions represented by the selected components.

The connection between these component choices and acceptable dispatch outcomes is established through numerical evaluation.
Fig.~\ref{fig:Compositions}(b) shows that an accepted candidate is obtained for every daily task, although the initial candidate is accepted only on Day~1.
The reported dispatch performance thus reflects successive candidate evaluations beyond the initial model proposal.
Furthermore, additional candidates can be tested after an accepted formulation has been found, while an earlier accepted candidate remains available for final selection.
This separates candidate acceptance from search termination, preserving a validated option during further exploration.

\begin{figure}[h]
\centering
\includegraphics[width=\columnwidth]{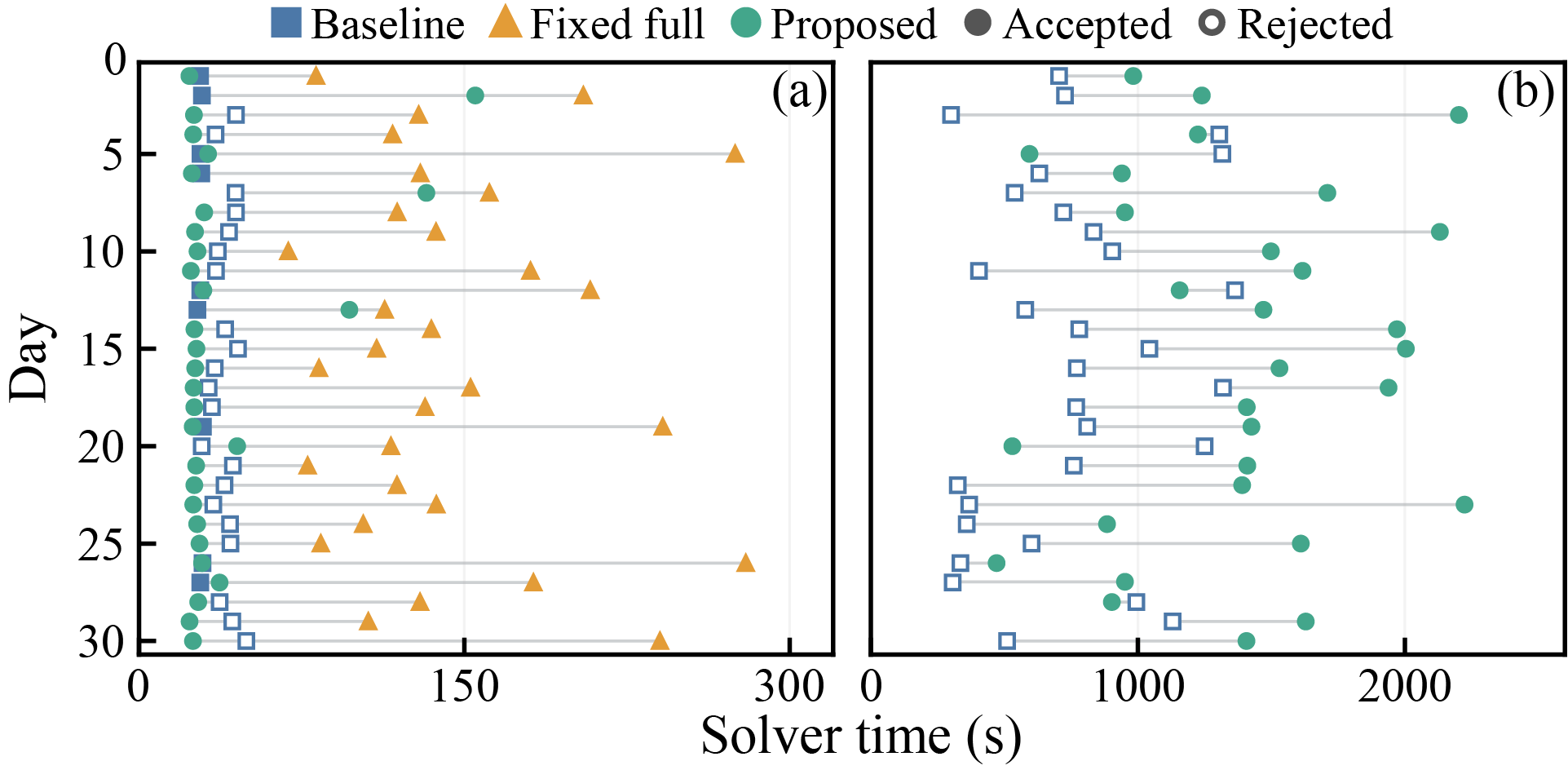}
\caption{Computational Performance over 30 days.
(a) T9D2: fixed simplified baseline, fixed full formulation, and proposed adaptive composition.
(b) T118D10: fixed simplified baseline and proposed adaptive composition. }
\label{fig:Speed}
\end{figure}

The computational implications of the component selections are examined through the daily solver times and validation outcomes in Fig.~\ref{fig:Speed}.
On T9D2, the proposed compositions and the fixed full reference are accepted on all 30 days, whereas the fixed simplified baseline is accepted on only nine days.
The proposed formulations require less solver time than the fixed full reference on every daily task.
The relatively small gains on Days~2,~7, and~13 coincide with the retention of full DistFlow in Fig.~\ref{fig:Compositions}(a), linking the computational comparison to the network representations selected for these tasks.

\begin{table*}[!b]
\caption{LLM-guided model revision on T9D2 (Day 20)}
\label{tab:t9d2_complete_llm_trace}
\centering
\footnotesize
\setlength{\tabcolsep}{3pt}
\renewcommand{\arraystretch}{1.05}

\begin{tabularx}{\textwidth}{@{}
c
>{\raggedright\arraybackslash}p{43mm}
>{\raggedright\arraybackslash}X
>{\raggedright\arraybackslash}p{24mm}
>{\raggedright\arraybackslash}p{31mm}
@{}}
\toprule
Round
& Validation feedback
& Composition decision
& TN / DN formulation
& Outcome \\
\midrule

1
& Not invoked: initial selection.
& Retain voltage/reactive-power and loss states
  without full nonlinear network equations.
& SOCP / SOCP
& Rejected; 1113 crossings;
  $V_{\max}=1.0601$; taps converged. \\

2
& After R1: reconsider TN/DN voltage fidelity;
  cause not localized.
& Test full TN/DN relations jointly because R1
  does not localize the approximation gap.
& ACOPF / Distflow
& Accepted; 204.30~s.
  Retain R2. \\

\addlinespace[2pt]

3
& Not invoked: R2 accepted.
& From R2, relax TN only to test whether full TN relations are necessary.
& SOCP ACOPF / Full Distflow
& Rejected; 218 crossings;
  $V_{\max}=1.0584$. \\

4
& After R3: preserve voltage/reactive-power coupling; voltage path not localized.
& From R2, relax DN only; retain full TN to isolate DN fidelity requirements.
& ACOPF / SOCP Distflow
& Accepted; 143.07~s.
  Retain R4. \\

5
& Not invoked: R4 accepted.
& From R4, test linear DN equations with fixed-factor losses to reduce structural burden.
& ACOPF / LINDistflow
  \newline (fixed loss)
& Rejected; 8 crossings;
  $V_{\min}=0.9473$. \\

6
& After R5: report undervoltage; request exact voltage checks and path attribution.
& From R4, retain DN current/loss states; replace SOC coupling by P/Q planes to test whether the cone is necessary.
& ACOPF / SOCP--LINDistflow
& Accepted; 60.01~s.
  Retain R6. \\

7
& Not invoked: R6 accepted.
& From R6, remove DN current/loss states to test whether a lossless DN is sufficient.
& ACOPF / LINDistflow
& Rejected; 8 crossings;
  $V_{\min}=0.9469$. \\

8
& After R7: undervoltage persists; retain voltage/reactive checks and path attribution.
& From R6, relax TN voltage-product equality; retain DN SOCP--LINDistflow to test remaining TN burden.
& SOCP / SOCP--LINDistflow
& Rejected; 705 crossings;
  $V_{\max}=1.0656$. \\

\midrule
\multicolumn{5}{@{}p{\textwidth}@{}}{%
\textbf{Final return:}
The execution limit is reached after R8; R6 (ACOPF / SOCP--LINDistflow) is returned, with a numerical execution time of 60.01~s.} \\
\bottomrule
\end{tabularx}

\par\vspace{2pt}
\begin{minipage}{\textwidth}
\footnotesize
\raggedright
\end{minipage}
\end{table*}

On T118D10, the comparison focuses on the proposed workflow and the fixed simplified baseline.
The fixed full reference is not included owing to the long solver times and computational difficulties associated with jointly solving the TN AC OPF and DN DistFlow models on the larger system.
The proposed workflow obtains accepted schedules for all 30 daily tasks, whereas none of the baseline schedules passes validation.
This improvement is accompanied by higher solver times on most days, highlighting the need to assess computational effort together with validation outcomes.

\subsection{Validation and Repair Performance}

Table~\ref{tab:t9d2_complete_llm_trace} examines how numerical feedback and subsequent trials produce the final model composition on T9D2 Day~20.
This case links the AC OPF/SOCP--LIN DistFlow combination reported in Fig.~\ref{fig:Compositions}(a) to the component revisions and acceptance decisions preceding its selection.

The initial candidate uses SOCP formulations for both the TN and DNs but fails the AC voltage check, with $V_{\max}=1.0601$~p.u.
The validation agent recommends reconsidering the voltage representations in network layers.
The composition agent therefore tests full formulations in both layers in R2, obtaining the first accepted candidate.
This accepted formulation provides a reference for subsequent simplification trials.

After acceptance, the composition agent continues to propose simpler alternatives, each subject to deterministic validation.
The two trials from R2 yield different outcomes: relaxing only the TN formulation in R3 is rejected, whereas retaining AC OPF and replacing full DistFlow with SOCP DistFlow in R4 is accepted.
These trials distinguish between alternative simplifications of the accepted formulation and establish R4 as the reference for further exploration.

Starting from R4, the linear DN formulation with fixed factor losses tested in R5 is rejected because of undervoltage.
Following this feedback, another trial from R4 retains the DN current and loss states while approximating the conic coupling using P/Q planes, producing the accepted SOCP--LIN DistFlow formulation in R6.
The reported numerical execution times of the accepted candidates decrease from 204.30~s in R2 to 143.07~s in R4 and 60.01~s in R6.

R7 and R8 subsequently test two separate simplifications from R6: removal of the DN current and loss states and relaxation of the TN voltage-product equality, respectively.
Both candidates fail validation, while R6 remains available for final selection.
When the execution limit is reached after R8, the workflow returns R6.
This trace illustrates how numerical feedback informs component revision, deterministic validation screens the resulting candidates, and candidate retention preserves an accepted formulation during further exploration.

\section{Conclusion}
This paper proposed TDMOpt, an LLM-driven multi-agent framework for adaptive model composition in coordinated TDM low-carbon dispatch.
By combining a layered executable component library with retrieved physics and component knowledge, the framework selects compatible models according to operating requirements.
Numerical feedback guides component revision, while deterministic validation and candidate retention support further exploration after an accepted formulation has been obtained.
Case studies on T9D2 and T118D10 showed that the framework obtains dispatch schedules satisfying the specified operational checks, with improved voltage compliance and a modest increase in operating cost relative to the fixed simplified baseline.
On T9D2, the adopted compositions also achieved substantial reductions in solver time relative to the fixed full reference while satisfying the same validation criteria.

Future work will expand the executable component library and integrate additional coordination and solution methods to support a broader range of TDM low-carbon dispatch tasks, including those involving forecast uncertainty.
Candidate exploration will also be improved by accounting for the computational cost of LLM inference, model assembly, numerical solution, and validation.

\end{document}